
\input jnl.tex
\def\fourth {{1\over 4}}
\doublespace
\vglue 0. truein
\title
{
Bound States Can Stabilize Electroweak Strings
}
\smallskip
\author
{Tanmay Vachaspati}
\affil
{
Tufts Institute of Cosmology, Department of Physics and Astronomy,
Tufts University, Medford, MA 02155.
}
\smallskip
\author
{Richard Watkins}
\affil
{
Physics Department, University of Michigan, Ann Arbor, MI 48109.
}

\abstract
\doublespace

We show that the electroweak $Z-$string can be stabilized by the
presence of bound states of a complex scalar field. We argue that
fermions coupled to the scalar field of the string can also
make the string stable and discuss the physical case where the
string is coupled to quarks and leptons. This stabilization mechanism
is expected to  work for other embedded defects and also for unstable
solutions such as the sphaleron.

\endtitlepage

It is now known that vortex solutions\refto{1} may be embedded in almost
any field theoretic model that exhibits spontaneous symmetry
breaking\refto{2}. In particular, two distinct vortex solutions are
known to be embedded in the standard
electroweak model\refto{2, 3, 4, 5}. These are called
the $\tau -$string and the $Z-$string in the literature. The
$\tau -$string is conjectured to be unstable for all values of
the parameters while the $Z-$string has been shown to be stable
only for ${\rm sin}^2\theta_W \approx 1$. The $Z-$string in the standard
electroweak model with ${\rm sin}^2\theta_W = 0.23$ is
unstable\refto{6} and remains so even at high temperatures
such as would be present in the early universe\refto{7}.

If the string solutions are indeed unstable under all circumstances,
their relevance to physical processes would probably be negligible.
However, in this letter we shall show that the strings can be stabilized
by the presence of other scalar or fermionic fields in the theory.
The idea behind this result is quite simple to understand, especially
if one is aware of the reason that permits the existence of
non-topological solitons\refto{8}. Suppose that we have a theory
in which the Higgs mechanism is responsible for generating the
mass of a certain scalar. Then, after the symmetry
breaking, the Higgs field gives the scalar a mass but the back-reaction
of the scalar field on the Higgs field is to try and prevent the Higgs
field from acquiring its vacuum-expectation-value (vev).
In other words, the scalar would rather live in a  region where the Higgs
field vanishes  since the mass of the scalar field is zero wherever
the Higgs field is zero. But the center of the string is precisely a region
where the Higgs field vanishes. Therefore the scalar likes to accumulate
on the string and tends to maintain the string configuration with
its region of vanishing Higgs
field - that is, the scalar adds to the stability
of the configuration.
Yet another way of stating this idea is that {\it the string is
a ``bag'' in which the scalar prefers to sit and, hence, hold
together}.

In what follows, we shall only consider the case of a scalar field
interacting with the electroweak $Z-$string.
To start with, we shall describe the effect
of scalar bound states
on semilocal strings\refto{9} where it is fairly clear
that the stability improves due to the bound state. This in itself shows
that the electroweak $Z-$string will become more stable when it has
scalar bound states since, after all, the $Z-$string is nothing but the
semilocal string when ${\rm sin}^2\theta_W = 1$. However, we go
further and explicitly examine the case of the $Z-$string with a
scalar bound state. Our results suggest that it may be possible
to get stable $Z-$strings
even when ${\rm sin}^2\theta_W =0.23$.

This does not immediately imply that stable $Z-$strings occur in the
standard electroweak model since there is no extra scalar field
in this model. However, the standard model does
contain leptons and quarks which will also have bound states on the string.
We expect the
arguments of the previous paragraphs to apply in this case too since,
once again, it is favorable for the fermion to sit in the string ``bag''
and to prevent the bag from decaying.
Whether the lepton and quark bound states are sufficient to stabilize the
$Z$-string is another story that needs detailed
investigation. We hope to undertake this task in the near future.

The Lagrangian that yields semilocal string solutions with an additional
complex scalar field is:
$$
L_{sl} = ( D^\mu \phi )^{\dag} ( D_\mu \phi ) +
         ( \partial^\mu \chi )^* ( \partial_\mu \chi )
         - \fourth F_{\mu \nu}^Z F_Z ^{\mu \nu}
         - V(\phi , \chi )
\eqno (1)
$$
where,
$$
V(\phi , \chi ) = \lambda _1 \left ( \phi ^{\dag} \phi -
                   {{\eta^2} \over 2} \right ) ^2 + \lambda _2 |\chi |^4
                  +2 \lambda _3 (\phi ^{\dag} \phi \pm m^2 )
                  \chi ^* \chi  \  .
\eqno (2)
$$
The field $\phi$ is a global $SU(2)$ doublet carrying a gauged $U(1)$
charge, while $\chi$ is a
single complex field. The covariant derivative is defined by,
$$
D_\mu = \partial _\mu + {i \over 2} \alpha Z_\mu \ .
\eqno (3)
$$

There are two approaches to finding solutions that describe a string
with a non-trivial $\chi$ configuration.
The first is that, for the negative sign in (2) and for some values of
the parameters, the string configuration together with $\chi =0$ is
unstable, and the stable ground state solution is one that has a
non-trivial $\chi$ condensate on the string\refto{10}. It may be
speculated that the presence of a condensate\refto{11} might
improve the stability of the string. Indeed we have checked that there
is an improvement in string stability due to a condensate but the
improvement is only marginal and is certainly not enough to stabilize
the string when ${\rm sin}^2 \theta_W = 0.23$. The second approach is to
consider the string in the presence of $\chi$ particles - that is,
the string with $\chi$ bound states. The $\chi$ particles
carry a conserved $U(1)$ global charge which is derived from the
conserved current
$$
j^\mu =
{{i} \over 2} ( \chi^* {\partial^\mu \chi} -
\chi {\partial^\mu \chi}^* ) \ .
\eqno (4)
$$
Hence, we consider a string in the presence of a certain amount of
global $U(1)$ charge. This, together with cylindrical symmetry,
leads to the following ansatz for $\chi$:
$$
\chi = e^{i\omega t} \psi (r) \ ,
\eqno (5)
$$
where $r$ is the cylindrical radial coordinate.
The charge per unit length along the $z-$direction in this configuration is:
$$
q = 2 \pi \omega \int dr \ r \psi ^2 \  .
\eqno (6)
$$

We will look at solutions of the equations of motion following from (1)
that consist of a semilocal string and a fixed amount of $U(1)$
charge. In accordance with the usual ansatz for the semilocal
string\refto{9}, we take,
$$
\phi = \pmatrix{0\cr f(r)e^{i\theta}\cr} \  ,
\ \ \ \  Z_\mu = - {{v(r)} \over r} {\hat e}_\theta  \  .
\eqno (7)
$$
It is now
convenient to rescale the fields and coordinates to make them dimensionless:
$$
P = {{\sqrt{2}} \over \eta} f , \ \ \ \  V = {\alpha \over 2} v ,
\ \ \ \  w = {{4 \sqrt{\lambda_3}} \over {\alpha \eta}} \psi , \ \ \ \
R = {{\alpha \eta} \over {2\sqrt{2}}} r \ .
\eqno (8)
$$
Then the equations of motion are:
$$
P'' + {{P'} \over R} - (1-V)^2 {P \over {R^2}} +
\beta (1-P^2 )P - w^2 P = 0 \ ,
\eqno (9)
$$
$$
V'' - {{V'} \over R} + 2(1-V) P^2 = 0 \ ,
\eqno (10)
$$
$$
w'' + {{w'} \over R} - \lambda w^3 - \gamma (P^2 - \delta ^2 ) w = 0
\eqno (11)
$$
where primes denote derivatives with respect to $R$.
The parameters entering these equations are defined by:
$$
\beta = {{8\lambda_1} \over {\alpha^2}} = {{m_H^2} \over {m_Z ^2}} \ ,
\ \ \ \  \lambda = {{\lambda _2} \over {\lambda_3}} \ , \ \ \ \
\gamma = {{8 \lambda_3} \over {\alpha^2}} = {{m_\chi ^2} \over {m_Z ^2}} \ ,
\ \ \ \  \delta^2 = {2 \over {\eta ^2}}
\left ( m^2 + {{\omega ^2} \over {2\lambda_3}} \right )  \ .
\eqno (12)
$$
Here $m_H$, $m_Z$ and $m_\chi$ denote the masses of the $\phi$, $Z$
and $\chi$ particles respectively. In addition to the equations (9), (10)
and (11), we also have the constraint that the rescaled
(dimensionless) charge is some fixed  non-zero constant. Therefore,
$$
\bar q \equiv { {\omega} \over {\eta \sqrt{\lambda_3}} }
\int dR \ R \ [w(R)]^2 = {\rm constant} \ .
\eqno (13)
$$
The boundary conditions on $P$, $V$ and $w$ are:
$P(0) = 0$, $P(\infty ) =1$, $V(0) = 0$, $V(\infty ) = 1$,
$w'(0) =0$ and $w(\infty ) = 0$.

So far we have been looking at the {\it unperturbed} string plus bound state
solution. Now we turn to the stability analysis.

The stability
analysis of the semilocal string\refto{12} can be
reduced to an analysis of the perturbation in the upper component
of $\phi$ alone. Even in the presence of a bound state, this
remains true since it is the upper component of $\phi$ which provides
a channel for the string to unwind on the vacuum manifold. If the upper
component of $\phi$ was forced to remain zero, the semilocal string
would be identical to the Nielsen-Olesen string which we know to be
topologically stable even in the presence of other fields. Hence, it
is sufficient to examine perturbations in $\phi_1$ - the rescaled
upper component of $\phi$. Furthermore, it is sufficient
to consider $\phi_1$ to be real and a function of the radial coordinate
alone\refto{6,13}.

The energy variation due to the perturbation $\phi_1$ is:
$$
\delta E = \eta^2 \pi \int dR R \left [
                                 {\phi_1 '} ^2 + M^2 (R) \phi_1 ^2
                          \right ] \ ,
\eqno (14)
$$
where,
$$
M^2 (R) = {{V^2} \over {R^2}} + \beta ( P^2 - 1 ) + w^2 \  .
\eqno (15)
$$
It is immediately obvious that the presence of a bound state improves
the stability of the semilocal string since the contribution to $M^2$
coming from $w$ is always positive. In the absence of
a bound state, we know that the semilocal string is stable only
for\refto{13, 14}
$0 \le \beta \le 1$. Hence, a bound state on the semilocal string
will stabilize the string for values of $\beta$ larger than 1.
A quantitative statement about the stability of the
``bound semilocal string'', however,
requires a numerical analysis since the bound state will also back-react
on the unperturbed string configuration.
Here, since we are primarily interested in the electroweak string,
we simply remark that our numerical analysis confirms that the
semilocal string can be stabilized for $\beta > 1$ if we include a
suitable bound state on the string.

The electroweak string will have the same field configuration as
the semilocal string, with all the gauge fields except the $Z$
set to zero.
To analyze its stability, we use the results
of Ref. \cite{6}, where it is shown that the stability issue reduces
to asking if there are any negative eigenvalues ($\Omega$)
to the Schr\"odinger equation:
$$
- {1 \over R} {{d\ } \over {dR}} \left (
   {R \over Q} {{d\zeta} \over {dR}} \right ) + U(R) \zeta = \Omega \zeta
\eqno (16)
$$
with
$$
U(R) = {1 \over Q} {{{P'} ^2} \over {P^2}} + {{2 S} \over {R^2 P^2}}
         + {1 \over R} {{d\ } \over {dR}} \left (
                {R \over Q} {{P'} \over P} \right )
\eqno (17)
$$
$$
Q = (1 - 2 cos^2 \theta_W V )^2 + 2 cos^2 \theta_W R^2 P^2
\eqno (18)
$$
$$
S = {{P^2} \over 2} - cos^2 \theta_W {{{V'} ^2} \over Q}
    + {R \over 2} {{d\ } \over {dR}} \left [
     {{V'} \over R} {{(1-2 cos^2 \theta_W V)} \over Q} \right ] \  .
\eqno (19)
$$
The boundary conditions on $\zeta$ are: $\zeta (0) = 1$,
$\zeta ' (0) = 0$ and $\zeta (\infty ) = 0$.

We have first found the unperturbed configuration using (9), (10)
and (11) subject to the constraint (13) by using numerical relaxation
techniques. Then we have solved (16) by a numerical shooting method
which allows us to check if $\Omega$ is positive or negative. In all
our numerical work we have taken $\beta = 0.40$
($m_H = 58 \ GeV$) - this satisfies the experimental constraint
$\beta > 0.38$ obtained by LEP. In Fig. 1 we show the
minimum value of ${\rm sin}^2 \theta_W$ that is required for a string
with a certain amount of charge to be stable in the case when
$\lambda = 0 = m$ and $\gamma = 1$. (The plot is not very sensitive
to the value of $\gamma$.)
This plot shows that the stability of the
electroweak $Z-$string greatly improves in the presence of bound
states and stability at lower values of ${\rm sin}^2 \theta_W$ may be
achieved by putting enough charge on the string. Although our
numerical analysis wsa not able to find a stable solution for
the physical case ${\rm sin}^2 \theta_W=0.23$, we feel that a more extensive
exploration of parameter space might result in a stable solution
in this case also. On the other hand, this issue is not very relevant
since the truly physical model does not contain an extra scalar field
like $\chi$.

The reader may wonder if the introduction of the field $\chi$ could
have introduced some new instability in the configuration. As was
shown in Ref. \cite{25} in the case of cylindrical non-topological
solitons, there is a possible instability in the distribution of
charge along the string. We have checked that this instability is also
present in the electroweak string with $\chi$ bound states at least
in the case when $m^2 = 0$. A little thought, however, reveals that
the instability is peculiar to our toy model and would be absent
in the physically realized model.
This is because $\chi$ and $\phi$, being spin 0 fields, lead to an attractive
force between the $\chi$ charges, so that the unperturbed linear distribution
of charge on the string is unstable to clumping up
into spherical distributions. In the physical model, however,
there are gauge fields (spin 1) present in the theory which would lead
to repulsive forces between the bound charges and would prevent
the clumping instability\refto{15}.

The most pertinent question at this juncture is if the standard electroweak
model also admits stable bound electroweak strings.
In this case one needs to look at fermionic bound states with the standard
number of quarks and leptons and with the experimentally determined
parameters.
The physical argument - in which we view the string as a bag - applies to
fermions also and so fermionic bound states will also improve the stability
of electroweak strings. One difference with the bosonic case is that
fermions obey the Pauli exclusion principle and so every additional
fermion that we put in the string bag must occupy a different quantum
state.  This makes it somewhat less energy efficient to pack
fermions onto the string\refto{16}.

There are some additional (technical) difficulties in investigating
the bound $Z-$string in a realistic setting. The first such
difficulty is that the quarks and leptons carry electromagnetic charge
and the electromagnetic field of the bound state must also be taken
into account in the stability analysis. A second difficulty is that the
stability analysis must necessarily include the neutrinos since these
couple to the perturbations of the Higgs field. Both these difficulties
promise to make the realistic stability analysis an Herculean task.

The stabilizing effect of bound states would be felt by other
embedded defects as well\refto{17}.
In particular, one may ask
if bound states can stabilize the electroweak $\tau -$string.
It would also be of some interest to study the effects
of bound states on embedded monopole configurations\refto{2}.
Given the rather
general stability analyses of monopoles\refto{18, 19}, it would
be worthwhile to see how bound states can fail to stabilize the
embedded monopole.
Another unstable configuration that might be stabilized by the presence
of bound states is the sphaleron\refto{5}.

Finally we would like to make one more comment that is relevant for
cosmology and the observational prospects for electroweak defects.
A {\it loop} of electroweak string has two distinct instabilities:
the first is the field-theoretic instability that we have discussed
above and the second is a dynamical instability against collapse.
In this letter we have argued that the presence of bound states
on the string can protect the string against
the field-theoretic instability. These bound states also add to
the energy density of the string without correspondingly adding to the
pressure.  This implies that the dynamics of bound strings should be
similar to that of wiggly strings\refto{20} or to current carrying
superconducting strings - depending on the nature of the charge on the
string. It is also known from previous
work that currents\refto{10} on the string can protect the loop against
dynamical collapse\refto{21, 22, 23, 24}.
These facts together suggest the possibility
that there are stable, static ring configurations
(also, vortons\refto{22}) in the standard electroweak model.
If the sphaleron is also stabilized by bound states, it would
imply additional particle-like solutions in the model.

\bigskip

\beginsection {\it{Acknowledgements:}}

We are grateful to Alex Vilenkin and Jose Wudka for discussions.
TV acknowledges the support of the NSF
and RW acknowledges the support of
NASA Grant No. NAGW-2802.


\references

\refis{3} T. Vachaspati, Phys. Rev. Lett. {\bf 68}, 1977 (1992);
 T. Vachaspati, Nucl. Phys. B, to be published.

\refis{9} T. Vachaspati and A. Ach\'ucarro, Phys. Rev. D {\bf 44},
3067 (1991).

\refis{1} H. B. Nielsen and P. Olesen, Nucl. Phys. B{\bf{61}}, 45 (1973).

\refis{13} M. Hindmarsh, Phys. Rev. Lett. {\bf 68}, 1263 (1992).

\refis{14} A. Ach\'ucarro, K. Kuijken, L. Perivolaropoulos and
T. Vachaspati, Nucl. Phys. B, to be published.

\refis{4} Y. Nambu, Nucl. Phys. B{\bf 130}, 505 (1977).

\refis{2} T. Vachaspati and M. Barriola, Phys. Rev. Lett. {\bf 69},
1867 (1992).

\refis{5} N. S. Manton, Phys. Rev. D{\bf 28}, 2019 (1983).

\refis{6} M. James, L. Perivolaropoulos and T. Vachaspati,
Phys. Rev. D, to be published.

\refis{7} R. Holman, S. Hsu, T. Vachaspati and R. Watkins,
``Metastable Strings in Realistic Models'', Fermilab Pub. 92/228-A (1992).

\refis{10} E. Witten, Nucl. Phy. B{\bf 249}, 557 (1985).

\refis{21} E. Copeland, M. Hindmarsh and N. Turok,
Phys. Rev. Lett. {\bf 58}, 1910 (1987).

\refis{23} C. Hill, H. Hodges and M. Turner, Phys. Rev. D{\bf 37},
263 (1988).

\refis{24} A. Babul, T. Piran and D. N. Spergel, Phys. Lett. B{\bf 202},
307 (1988).

\refis{22} R. L. Davis and E. P. S. Shellard, Nucl. Phys. B{\bf 323},
189 (1989);
R. L. Davis, Phys. Lett. B{\bf 207}, 404 (1988); Phys.
Lett. B{\bf 209}, 485 (1988);
Phys. Rev. D{\bf 38}, 3722 (1980).

\refis{12} M. Hindmarsh, `` Semilocal Topological Defects'',
DAMTP preprint (1992).

\refis{8} T. D. Lee, in Proceedings of the Conference on Extended
Systems in Field Theory [Phys. Rep. {\bf 23C}, 254 (1976)].

\refis{18} R. Brandt and F. Neri, Nucl. Phys. B {\bf 161}, 253 (1979).

\refis{19} S. Coleman, Erice lectures 1981, ``The Magnetic Monopole
Fifty Years Later''.

\refis{11} Note the distinction between ``condensate'' and
``bound state''. A condensate is the ground state configuration of
$\chi$ in the background of the string while a bound state requires
the presence of $\chi$ particles.

\refis{17} The simplest example of an embedded defect is a domain
wall in a global $U(1)$ model.

\refis{25} E. Copeland, R. Kolb and K. Lee, Phys. Rev. D{\bf 38},
3023 (1988).

\refis{20} B. Carter, Phys. Rev. D{\bf 41}, 3886 (1990); A. Vilenkin,
Phys. Rev. D{\bf 41}, 3038 (1990).

\refis{15} Gauged non-topological solitons have
been studied in: K. Lee, J.A. Stein-Schabes, R. Watkins, and L.
Widrow, Phys. Rev. D{\bf 39}, 1665 (1989).

\refis{16} In both the bosonic and fermionic cases, the addition of
more charge is expected to lead to diminishing returns in improved
stability. This is
because the string ``swells'' as we increase the charge and this costs
energy.


\endreferences

\vfill
\eject

\head{Figure Captions}

1. The value of ${\rm sin}^2 \theta_W$ above which strings with a certain
amount of charge $\bar q$ are stable plotted versus $\bar q$ when
$\beta = 0.4$, $m=0=\lambda$ and $\gamma = 1$. For the largest value
of $\bar q$ that we have considered, the value of ${\rm sin}^2 \theta_W$
goes down to 0.46.

\endjnl
\end